\documentclass[conference]{IEEEtran}
\IEEEoverridecommandlockouts
\usepackage{cite}
\usepackage{amsmath,amssymb,amsfonts}
\usepackage{algorithmic}
\usepackage{graphicx}
\usepackage{textcomp}
\usepackage{xcolor}
\def\BibTeX{{\rm B\kern-.05em{\sc i\kern-.025em b}\kern-.08em
    T\kern-.1667em\lower.7ex\hbox{E}\kern-.125emX}}
    \usepackage{wrapfig}

\usepackage{fancyhdr} 
\fancyheadoffset{0cm}

\fancypagestyle{plain}{%
   \fancyhf{}%
   \fancyfoot[R]{\thepage}%
}
\begin{document}

\title{Learning-Driven Wireless Communications,\\ towards 6G\\
\thanks{978-1-7281-2138-3/19/\$31.00 \textcopyright 2019 IEEE}
}

\author{\IEEEauthorblockN{Md. Jalil Piran }
\IEEEauthorblockA{\textit{Computer Engineering Department} \\
\textit{Sejong University}\\
Seoul, South Korea \\
piran@sejong.ac.kr }
\and
\IEEEauthorblockN{Doug Young Suh}
\IEEEauthorblockA{\textit{Electronic Engineering Department} \\
\textit{Kyung Hee University}\\
Yongin, South Korea \\
suh@khu.ac.kr  }
}

\maketitle 
\thispagestyle{plain}
\pagestyle{plain}

\begin{abstract}
The fifth generation (5G) of wireless communication is in its infancy, and its evolving versions will be launched over the coming years. However, according to exposing the inherent constraints of 5G and the emerging applications and services with stringent requirements e.g. latency, energy/bit, traffic capacity, peak data rate, and reliability, telecom researchers are turning their attention to conceptualize the next generation of wireless communications, i.e. 6G. In this paper, we investigate 6G challenges, requirements, and trends. Furthermore, we discuss how artificial intelligence (AI) techniques can contribute to 6G. Based on the requirements and solutions, we identify some new fascinating services and use-cases of 6G, which can not be supported by 5G appropriately. Moreover, we explain some research directions that lead to the successful conceptualization and implementation of 6G.
\end{abstract}


\section{Introduction}
The future mobile trend is showing that the number of both unique subscribers as well as mobile connections will increase over the coming years. For instance, one forecast shows that there will be 6.1 billion unique subscribers and 9.1 billion mobile connections as of 2025 \cite{1}. The first standard of the fifth generation, referred as 5G NR (new radio), was released in June 2018 \cite{2} and launched at the beginning of April 2019 from South Korea. It is predicted that it will capture a 15\% share of the global market until 2025 \cite{1}.

\begin{figure}[t]
  \centering
    \includegraphics[width=8.9cm, height=5cm]{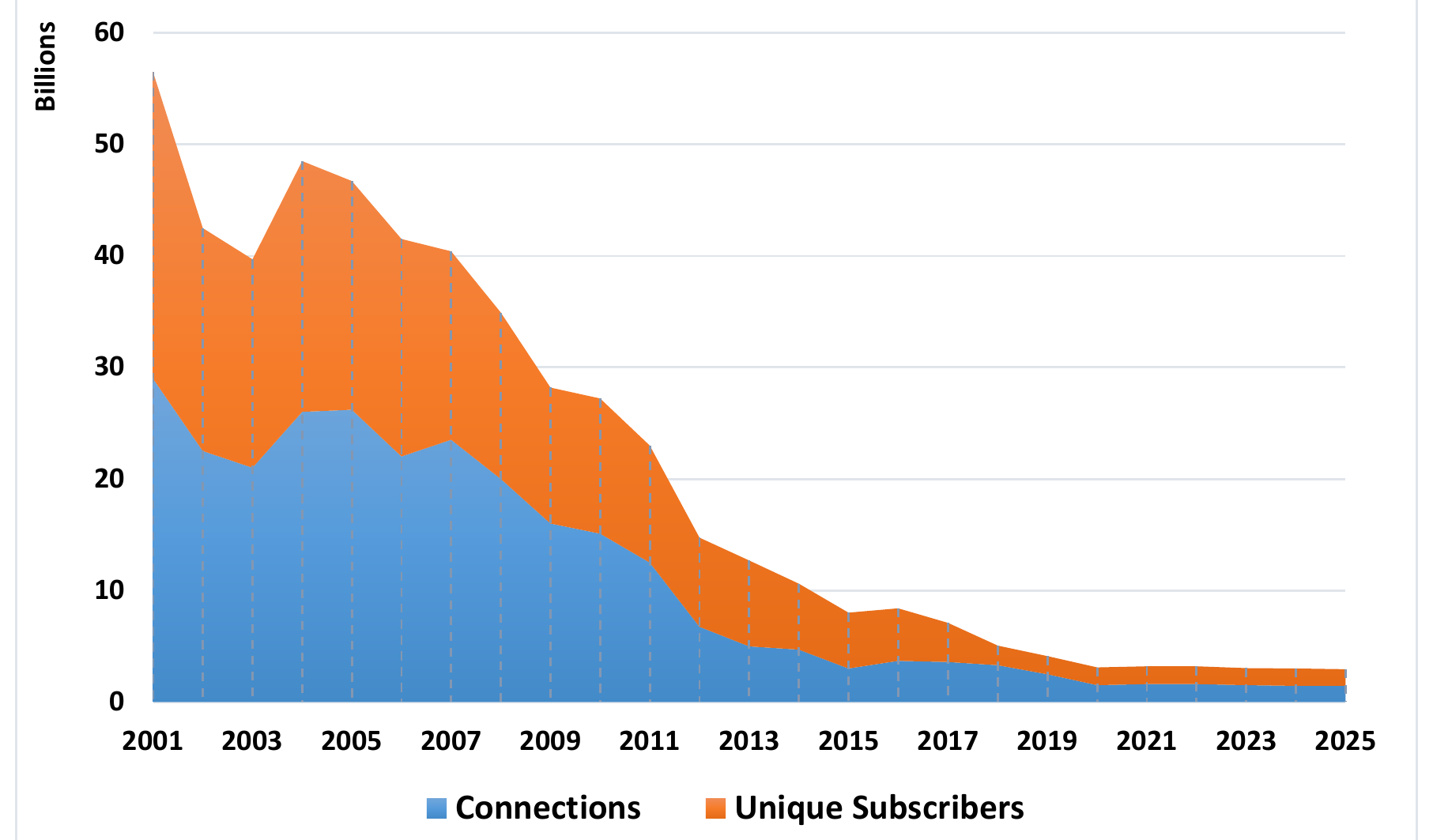}
  \caption{Growth rate of mobile industry.}
  \label{fig1}
\end{figure}

However, the growth rate of the number of unique subscribers as well as average revenue per user (ARPU) is decreasing in an unpleasant manner. For example, the growth rate of the number of subscribers was 20.5\% and 5\% in 2005 and 2015 respectively, while it will experience a lower rate of 1.45 as of 2025 \cite{1}, as shown in Figure \ref{fig1}. Such a salient reduction in the growth rate is considered as a crucial crisis for mobile network operators (MNOs) and asks for more attention from both academia and industry parties to save the industry by introducing some new charming and exciting services rather than only broadband services. 

According to the abovementioned statistics, and whereas the boundaries between mobile industry and the other verticals are blurred and while the MNOs are scrambling to monetize 5G, it seems to be the right time to turn the research attention beyond 5G in order to explore new business opportunities in the ever tremendous changing competitive battlefield.  

One of the promising solutions can be the Internet of Everything (IoE) to compensate those losses as the number and the growth rate of IoE devices and connections are experiencing a magnificent increasing ratio. As shown in Figure 2, an increased rate of 2.5 times for consumer users and 4.7 times industrial uses is predicted. IoE inherits of 5G requirements and performance, which is known also as massive machine type communications (mMTC). 

Moreover, the users expect some fascinating services beyond broadband applications that the previous generations were not capable to offer, such as extended reality (XR) (virtual reality (VR), mixed reality (MR), augmented reality (AR)), super Hi-vision (8K) video streaming, holographic communications, holoportation/telepresence, tele-surgery, haptics, unmanned aerial vehicles (UAV), connected autonomous vehicles, etc. 

As an instance, consider K-city, which is a fully autonomous driving smart city in South Korea. K-city is going to host thousands of self-driving vehicles in the near future. Such vehicles need super-precision positioning, e.g. close to centimeter accuracy, context-awareness to adapt to the changes in the environment in which they are running including the other vehicles, cyclist, pedestrians, as well as close to zero perceived latency to make feasible coordination and collaboration among the vehicles. Thus, it is necessary to minimize the journey times, fuel consumption, cost, and prevent possible collision and accidents.  

The abovementioned services demand stringent requirements such as close to zero latency, scalability, connection density, peak data rate and user-experienced data rate, traffic volume density, reliability, mobility, that the policy-driven 5G technologies are debatable to serve them appropriately. Among many, one reason is that early 5G rollouts mostly centered on sub-6GHz bands, particularly for mobility support.

Hence, as already happened for the previous mobile generations, the research phase and early-on visions start a decade prior to standardization, industrial implementation, and massive commercialization of a generation. Therefore, while the MNOs are scrambling to monetize 5G, academia starts research activities to conceptualize 6G.

In this paper, we explain the stringent requirements and trends of the next generation of mobile networks. Then, we describe how AI and ML techniques can help to address those challenges. Particularly, we explain how ML techniques can be employed in device processing in order to enable new services as well as minimize the network load. Then, we will discuss how different ML algorithms are helpful for network management in different layers. According to the identified requirements and solutions, we then introduce some envisioned services and use-cases that will be offered by 6G. Finally, some research directions towards 6G will be discussed.

The rest of the paper is organized as follows. Section II, presents 6G requirements and trends. In Section III, we discuss how AI techniques can be employed in order to enable intelligent network management and device processing. We provide some new services and use-cases that will be supported by 6G. Section V offers insight into future research directions. Finally, Section VI draws the conclusions.

\section{6G Requirements And Trends}
Based on the defined issues that the current generations of mobile networks struggling with, the mobile industry should be moved from the traditional strategy to some new ones including operation in shared spectrum bands, inter-operator spectrum sharing, indoor small cell networks, large number of local network operators, leasing network slice on demand, and many more. Thus, the requirements of 6G compared to 5G are more stringent as shown in Figure \ref{fig3}. In what follows, we discuss some of the most important requirements and trends of the next generation of mobile networks. 
\begin{figure}[t]
  \centering
    \includegraphics[width=8.9cm, height=5cm]{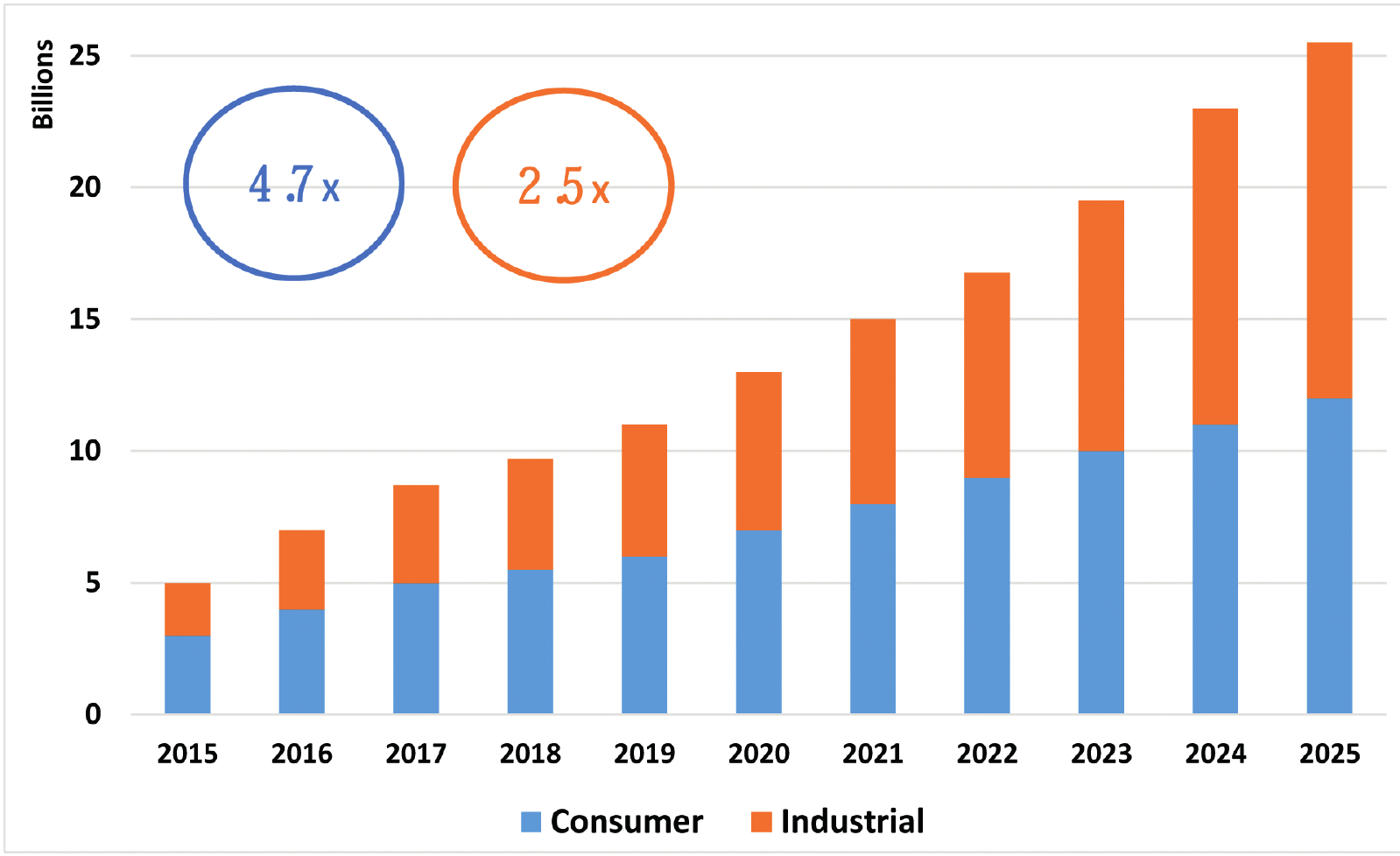}
  \caption{Growth rate of IoE.}
  \label{fig2}
\end{figure}
\begin{itemize}
\item \textit{Broad frequency bands}: Although sub-6 (3GHz to 6GHz) and mmWave bands (e.g. 24GHz, 28GHz, 37GHz, 39 GHz, 47 GHz) are supposed to be utilized by NR in order to support bandwidth-hungry services, the future networks ask for higher spectrum technologies, e.g. 73 GHz ~140GHz, 1THz~3THz or even optical communications. 

\item 	\textit{Opportunistic datarate}: As one of the most important trends, mobile traffic will experience a magnificent increase in the coming years. By the introduction of new services such as supper immersive multimedia, XR, holographic 3D video communications, 8K, 4K, etc., it is predicted that the dominant, i.e. 78\%, of mobile traffic will be video uploading/downloading \cite{1}. As an example, Samsung already started a project to make the Exynos9820 chip for smartphones, which enables the future smartphone to record 8K video content. Moreover, digital cinema (e.g. 64K) is also another mega-trend that experts are targeting. Such kind of services needs a very high and opportunistic peak data rate. 360° VR/AR services demand a bitrate of up to 500Mbps, while holographic communications need a bandwidth of 4 to10Tbps.

\item	\textit{Opportunistic latency}: New services in 6G are expected to have latency close to zero, e.g., less than 1ms end-to-end delay. For example, a 20ms latency in a networked AR application is very annoying. Similarly, the other services can tolerate only very low millisecond latency. As an example, telepresence needs sub-millisecond latency only.

\item	\textit{mMTC}: As predicted by Statista 2019, there will be 50 billion connected devices to IoE by 2030, which means a huge number of connections per km$^2$. In future mMTC, a.k.a. massive IoE (mIoE), machines play the roles that were done by human in the previous generations, e.g. machine-centric instead of human-centric. The main issues in mMTC that 6G will face is scalable and efficient connectivity for the huge number of connected devices, quality of service (QoS) provisioning, handling highly dynamic and sporadic mMTC traffic, huge signaling overhead and network congestion. Furthermore, 6G is required to enable wide area coverage as well as deep indoor penetration while minimizing the cost and power consumption. 

\item	\textit{Super-reliable MTC (sMTC)}: Most of the 6G uses-cases are not only bandwidth-hungry and delay-sensitive but also demand fast connectivity and high availability i.e.  \%99.999. Therefore, the future network must care about availability, latency, as well as reliability, which is known as sMTC. 

\item	\textit{Self-X network}: The future network must be more flexible and robust, which will be out of human ability to manage. Modern intelligent and adaptive ML techniques are adopted to support 6G network autonomy and to capture insights and comprehension on the environment. Hence, the future network will be able to perform the required functionalities such as self-learning, self-reconfiguration, self-optimization, self-healing,  self-organization, self-aggregation, and self-protection, without any human intervention.

\item	\textit{Super-precision operating}: Precision in a very high level and guaranteed, e.g. absolute deliver times, is of vital importance for many services. The traditional statistical multiplexing is not sufficient for such high-precision services. Therefore, some new function components are required including user-network interface, reservation signaling, new forwarding paradigm, intrinsic and self-monitoring operation, administration, and management for network configuration. Tele-surgery, intelligent transportation system, are some instance of those services.  

\item	\textit{Super-precision positioning}: The current localization systems uses signal level and travel time as well as satellites (e.g. Augmented-Global Positioning System (A-GPS), Glonass, and Galileo) to estimate the position. However, they struggle error even in order of meters, which make them inappropriate for most new services. The emerging use-cases such as tele-surgery, tactile internet, needs high precision positioning even with sub-millimeter accuracy.

\item	\textit{Scalability}: It is predicted that IoE connects billions of devices ranging from high-end computers to sensors, actuators, smartphones, tablet, wearable, home appliances, vehicles, etc., to the Internet. Such connected devices produces a sheer amount of data. The data are meaningful when using some technologies, the relevant hidden knowledge could be discovered. Therefore, AI techniques and algorithms are employed in order to analyze data and mine knowledge. AI-empowered devices, including user equipment (UE) and edge devices in IoE are able to analysis, summarize data, and discover knowledge prior to transmitting the raw data. This results in transforming a huge amount of data to a suitable amount and thus improves bandwidth utilization reliability and energy efficiency as well as decreases traffic load of the network and latency. 

\item	\textit{Supper energy efficiency}: The 6G devices need much more energy compared to the previous generations as they are supposed to operate in higher frequency bands. Therefore, power consumption and energy efficiency is a critical challenge that need to be considered. However, with the development of new techniques we will see a level of supper energy efficiency and even battery-free IoE devices, e.g. energy harvesting in building automation and smart homes. 

\item	\textit{Connectivity in 3D coverage}: The future networks will be expanded beyond 2D to cover oceans, atmospheric and space in order to support incorporating terrestrial and aerial devices. Such type of networks will support various applications including underwater acoustic ad hoc and sensor networks, weather forecasting, climate monitoring, etc.

\item	\textit{Integration with satellites}: The future 6G communications will employ satellite communication technologies to provide global coverage. 6G will integrate telecommunication satellites, earth imaging satellites and navigation satellites in order to provide localization services, broadcast and Internet connectivity, as well as weather information to cellular users. One example can be providing high-speed Internet to fast trains and airplanes. 

\item	\textit{Software-defined networking (SDN)}: Dynamic and programmatically efficient network configuration can play a major role in network management for future 6G. Network function virtualization (NFV) enables the consolidation of network instruments onto big servers located at data centers, distributed network devices, or even at end-user premises. Moreover, network slicing offers a cognitive and dynamic network framework on-demand, which can support several virtual networks on top of shared physical infrastructure. 
\end{itemize}
In the next section, we will discuss how AI and ML algorithms can contribute to the evolution of mobile generations. 
\begin{figure}[t]
  \centering
    \includegraphics[width=0.5\textwidth]{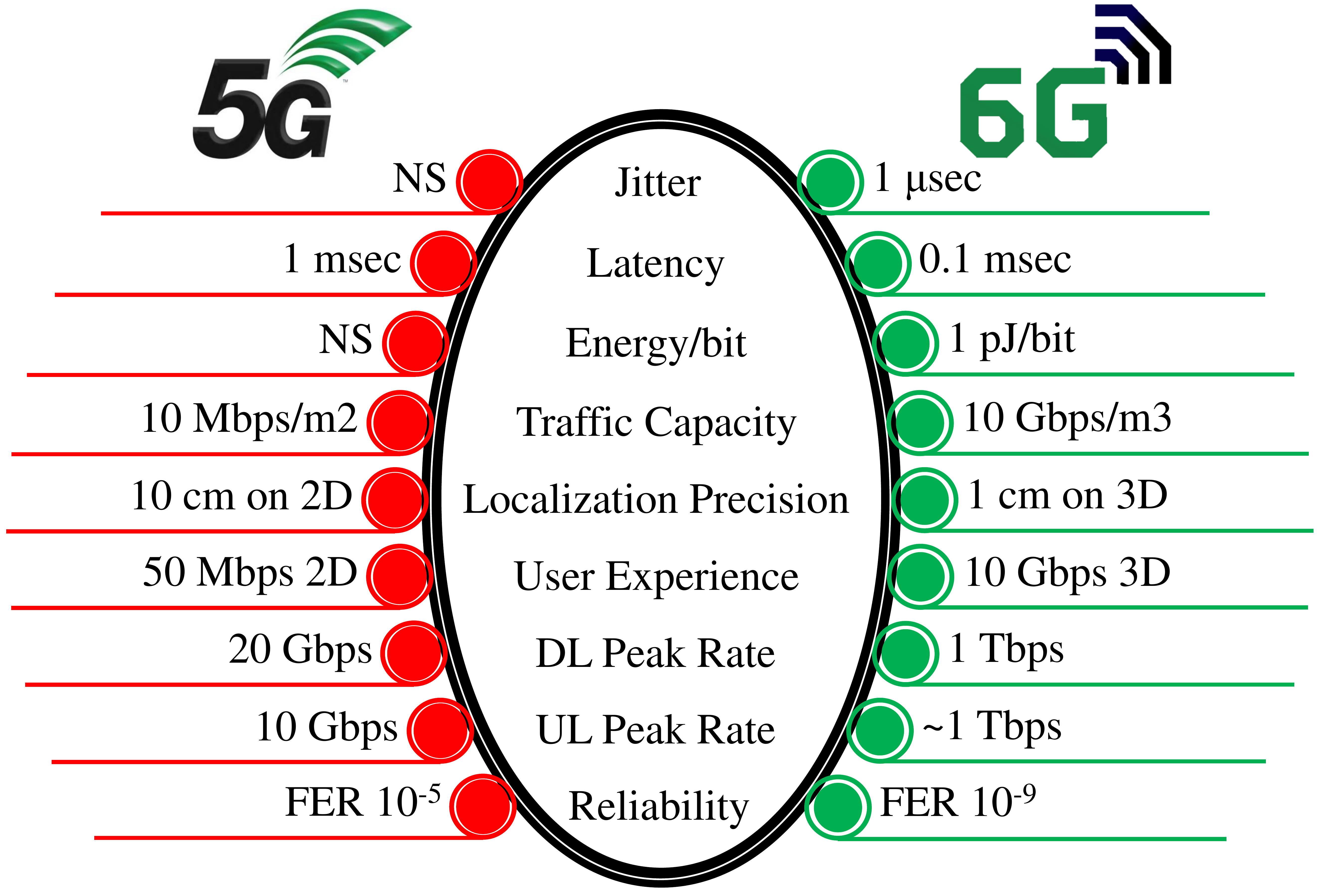}
  \caption{The requirements of 5G and 6G.}
  \label{fig3}
\end{figure}

\section{Learning-Driven Wireless Communications}
6G will be a transformative and revolutionized generation of wireless communications in a variety of aspects including context-awareness, ubiquitous, reconfigurable, and intelligent in both sectors i.e. IoE and mobile devices and network infrastructure. Thus, future networks will be too complex for human operators. Therefore, in order to achieve such capabilities, an AI-empowered and learning-driven network with the help of big data will be able to facilitate the users with unique capabilities including learning, reasoning, and decision-making to manage a network with zero-touch. AI is a big trend toward 6G for network resource management, network planning, and optimization, as well as failure detection and analysis. As a result, a combination of 6G-enabled IoE and AI will be much more captivating whereas IoE `supplies data' and AI `analysis data' and `mines knowledge'.

In the following subsections, we investigate how AI can be utilized in both network management and device processing in order to enhance operational reliability, real-time predictions, as well as increased security. 

\subsection{AI-powered Network Management}
AI improves network management in different aspects including monitoring, processing, and decision-making. Machine learning (ML) as a form of AI involves teaching the machines by making the data-driven decision (thinking) to perform tasks and functionalities (acting) independently without human intervention. ML has great potential in supporting big data analytic, efficient parameter estimation, and interactive decision-making. 

Subsequently, ML algorithms would make breakthroughs in the networking and communications fields, at both device and network levels, by enabling self-x networking (e.g. self-learning, self-reconfiguration, self-optimization, self-healing,  self-organization, self-aggregation, and self-protection). For example, ML can be used to address the access congestion in IoE \cite{3}, intelligence to the PHY to empower smart estimation of parameters, interference mitigation, and resource management \cite{4}, DL for channel estimation and symbol detection \cite{5}, DL for dynamic radio resource allocation for V2V and V2X \cite{6}.

In the following, we will explain how ML algorithms including supervised learning, unsupervised learning, and reinforcement learning, can be utilized in different network layers \cite{10}. 

\begin{itemize}
\item \textit{Supervised learning}: is a learning function that maps an input to an output based on the labeled training data, e.g. input-output pairs. There are many techniques developed under supervised learning that can be applied for network management including independent component analysis, locally linear embedding, principle component analysis, isometric mapping, K-means clustering, and hierarchical clustering.
Such algorithms are applicable in wireless networks in different layers. In the physical layer, we may utilize the above techniques for channel equalization, channel decoding, predicting path-loss and shadowing, channel states estimation, beamforming, adaptive signal processing, etc.
Supervised learning techniques are applicable in the network layer for caching, traffic classification, anomaly detection, throughput optimization, delay mitigation, etc.

\item \textit{Unsupervised learning}is a type of self-organized learning that is used to discover undefined patterns in a dataset without pre-determining labels. Various techniques have been developed for unsupervised learning including k-nearest neighbors, neural networks, decision tree and random forest, Bayesian learning, linear / logic regression, support vector machine.
In terms of network management, the above techniques can be used in network layer for routing, traffic control, parameter prediction, Resource allocations, RAT selection Handover and mobility management, Network slice selection and allocation, etc.
In physical layer unsupervised learning can be used for channel-aware feature-extraction, optimal modulation, interference cancelation, Channel estimation, MIMO precoding, node association, beam switching, etc.

\item \textit{Reinforcement learning} is about sequential decision-making, in which the agent ought to take actions in an environment with the goal of maximizing some notion of cumulative reward. Some of well-known techniques developed for network management are multi-arm bandit, temporal-difference learning, Markov decision process, SARSA learning, Q-learning, and Monte Carlo Method. These techniques can be applied in the application layer for proactive caching, data offloading, error prediction, and data rate allocation in network slicing. In the network layer, reinforcement learning can be used for multi-objective routing, packet scheduling, security, traffic prediction and classification, etc. While in physical layer, reinforcement-learning cab be employed for link preservation, channel tracking, on-demand beamforming, energy harvesting, modulation selection, radio identification, etc. 
\end{itemize}

\begin{figure}[t]
  \centering
    \includegraphics[width=7cm, height=4cm]{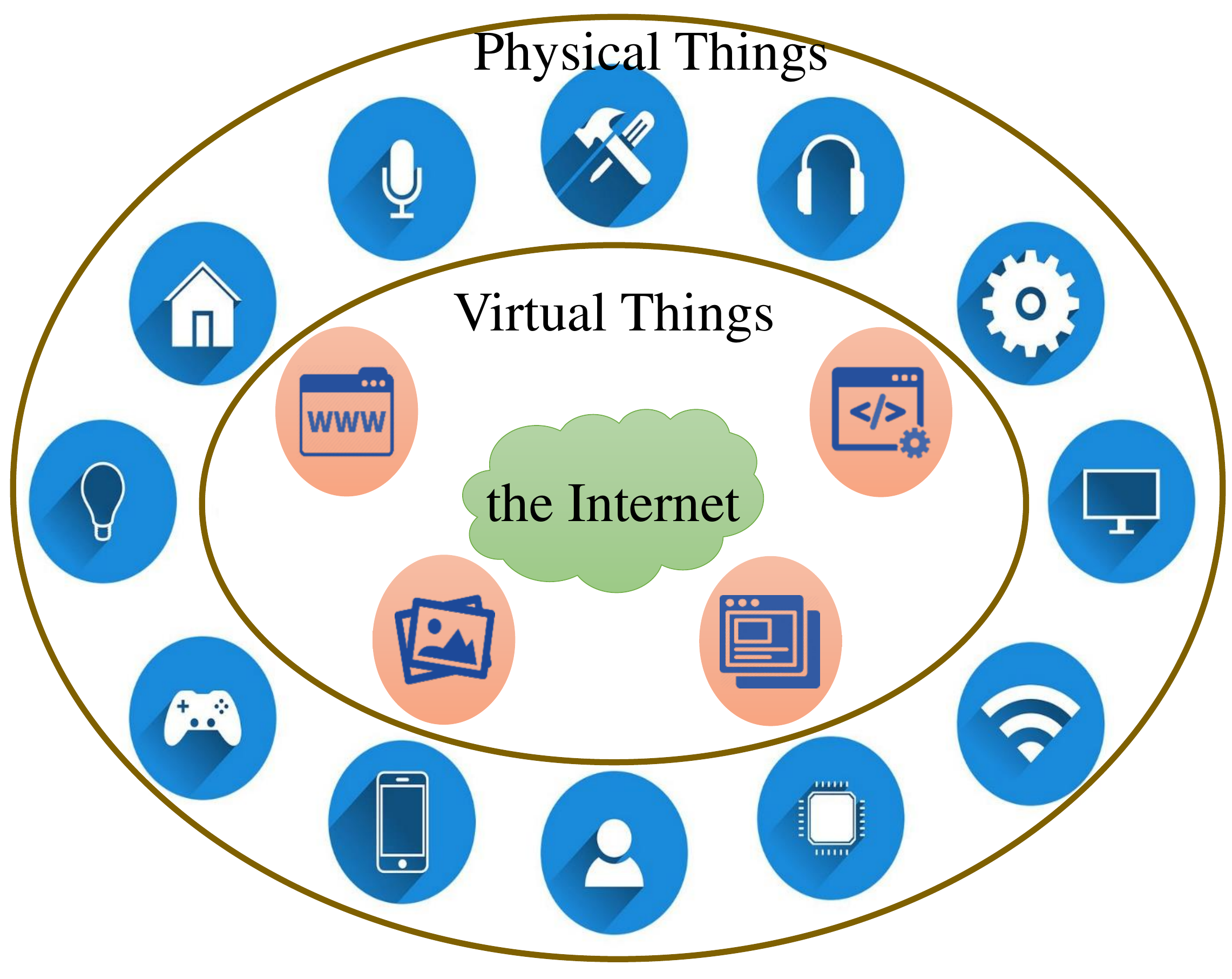}
  \caption{IoE is composed of physical and virtual things.}
  \label{fig4}
\end{figure}
\subsection{AI-powered Device Processing}
Recently, the trend of AI-powered devices is gathering momentum with the introduction of on-device Edge AI, where AI processing is transferred from the centralized computing facilities, e.g. cloud, to every terminal in the network, e.g. IoE devices, smartphone, BSs, APs, etc. Such smart and cloud-independent devices, by employing AI techniques in both hardware and software, would have a great impact on retail, healthcare, automotive, robotics, smartphones, and many more. A forecast carried by Strategy Analytic shows that 80\% of all smartphones would employ different AI techniques by 2023 \cite{7}. Hence, the vendors are trying to deliver an AI-based open ecosystem across all the devices and services to build the IoE, including both physical and virtual things, as shown in Figure \ref{fig4}. As a result, 6G-enabled IoE applications and services will be fascinating, much more than those services that IoE and AI provide independently. 

The most recent mobile phones are offering some new AI-based services. For example, new phones use AI techniques in their camera in order to increase the photo quality and look good aesthetically. AI-based cameras also take a photo based on some specific actions like raising a hand or pulling stupid faces, e.g. Photobooth.  Melaud is an AI-based smart earphone, control music based on body signals, e.g. heart rate and movement, while exercising. Bixby, Siri, Alexa, and Google Assistant are AI-powered intelligent interfaces, which can compose text messages from dictation via the voice app, identify landmarks, and translate foreign languages using its camera, shop online, operate home appliances, or provide reminders and recommendations based on the user’s schedule. Google Call screener answers calls verbally on the user and communicates with the caller, and then, summarizes the conversation in text format.

By developing ML algorithms and techniques, it is predicted that the future devices will be more smart. The followings are some predicted services that will be offered by mobile phone at the 6G era.

\begin{itemize}
\item An AI-based smart device using Neural Networks will be able to learn from the data that the user creates. 

\item Energy consumption efficiency is considered as one of the challenges for future mobile devices, which are going to operate on higher frequency bands, having very high resolutions screens, and the other energy-hungry applications. The future smart battery using ML techniques enables the devices to prioritze battery power for different applications and services.

\item	Neural processing engines are used on the chipsets to handle trillions of operations per second. On-device AR is getting help from such kind of powerful engines.

\item	Smart and cloud-independent devices are able to process the collected data and discover the desired knowledge close to the birthplace of data and hence reducing the network traffic volume as well as bandwidth consumption. when seeking knowledge about streets and highways traffic, crowded stadiums, malls, and subway stations, etc.

\item	Computational photography utilizes AI and computer vision for acquiring, processing, analyzing, and understanding digital photos to discover multi-dimensional knowledge from the physical environment.

\item	Samsung will release Exynos9820 chip that, by utilizing AI, offers a significant speed boost and improved energy efficiency. The chip enables the future smartphones to record 8K video at 30fps or 4K video at 150fps. 
\end{itemize}

\section{6G Envisioned Services and Use-cases}
\begin{figure}[t]
  \centering
    \includegraphics[width=8.9cm, height=7cm]{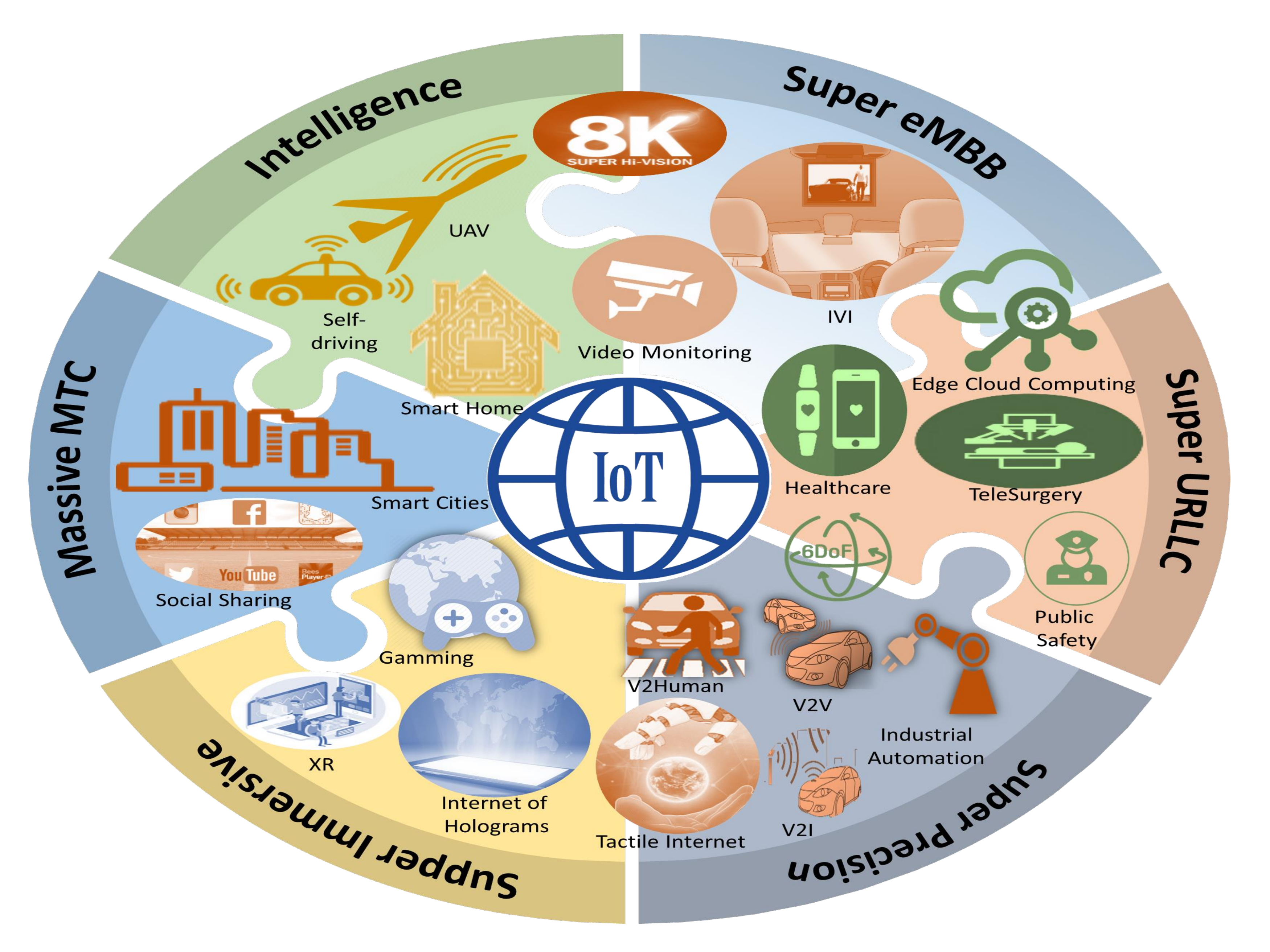}
  \caption{The envisioned services and use-cases for 6G.}
 \label{fig5}
\end{figure}

6G is predicted to be transformative in terms of connectivity and will enable not only human-to-human and human-to-machine, but also machine-to-machine in an intelligent way. Many new and emerging services and use-cases will be introduced in the 6G era, as shown in Figure \ref{fig5}.

The following are some services and use-cases, which cannot be served by the current technologies:

\begin{itemize}
\item \textit{Multisensory XR applications}: VR/MR/AR using various type of sensors collect data regarding orientation, acceleration, location, temperature, and audiovisual in order to provide an immersive experience. Such kind of applications demands very high data rate and reliability as well as close to zero perceived latency whereas they serendipitously collaborate. XR in high resolution, e.g. 4K and 8K, will be used in different applications including broadcasting, education, training, manufacturing, advertisement, medical, automobile, gaming, workspace communications, entertainment, etc.

\item	\textit{Internet of holograms}: holographic applications, e.g. holoportation and telepresence, enables true immersive communication in 5D of human sense information (including sight, hearing, touch, smell, and taste). Hence, near-real personal communication with digital holographic presence will be realized by the users. Moreover, by immersive live models, a combination of the environment with multiple digital avatars from different sources will be provided. Such kind of communications is infeasible using 5G whereas they demand even higher than 1Tbps data rate. For example, holographic telepresence of 77x20inch human requires 4.62Tbps of data rate. 

\item	\textit{In-Vehicle Infotainment (IVI)}: one of the exciting services that will be offer by the future networks is delivering high-quality services to the drivers and passenger on the go, such as in-vehicle ultra-high-quality terrestrial TV broadcasting. IVI enables the driver not only receive TV channels, but also secure firmware updates, map updates via terrestrial digital broadcasting facilities. This type of services demand very high bandwidth as well as high mobility support.

\item	\textit{High precision industry}: the future manufactures are too complex, which is out of human control, e.g. zero-touch. Such kind of systems requires reliability up to order of $10^-9$ and extremely low latency in the order of 0.1 to 1ms. 

\item	\textit{Tactile Internet (TI)}: the next evolution of IoE is TI, encompassing human-to-machine and machine-to-machine communications in real-time. Tactile applications such as haptics, requires super sRLLC, super-reliabile MTC, and high security. Interactive remote experiences often require latency 40ms while feedback below 5ms will enable novel uses of multi-sensory remote tactile control. TI covers a wide range of applications including, remote mining in high-risk areas, remote inspection, maintenance and repair of everything from manufacture to airplane, tele-robotic, etc.

\item	\textit{Smart city}: smart urban application including smart transportation, smart grid (SG), urban infrastructure, resident living environment, transportation management, medical treatment, shopping, security assurance, intelligent transportation system (ITS), intelligent medical diagnosis (IMD), shopping recommender system (SRS), which need a pervasive sensing and intelligent decision makers and actuators. 

\item\textit{Smart home}: the future intelligent homes will be able to provide users meticulous services, including energy management, patient assistance, real-time product labeling, and subscribing. 

\item	\textit{Six degrees of freedom (6 DoF) }: immersive content is considered as the next generation video with spatial movement for more immersive experiences that enables users to move freely. Such kind of services demands a tradeoff between throughput and latency that is 5 ms of latency and up to 5Gbps throughout.

\item	\textit{Social sharing at crowded venues}: event sharing are used in stadiums and other crowded areas, and massive simultaneous content is occasionally uploaded through social media. Such events may require up to 12.5 Tbps/km2 upload capacity.  

\item	\textit{e-Health}: new evolution of healthcare applications will be introduced by 6G with real-time tactile feedback, continuous connection availability, super low-latency data deliver, super high reliability, and high mobility support. 

\item	And many more cannot be served by 5G, including super Hi-vision (8K) or ultra-high definition (UHD) video streaming, connected robotics and autonomous systems, self-driving vehicles, UAV, wireless brain-machine interactions, blockchain and distributed ledger technologies, robotic autonomous drone deliveries.
\end{itemize}

\section{Research Directions}
The future 6G will not only concentrate on the pure communication field, but it will be interoperable between different but related fields including electronics and materials, wireless communication, and computer science and engineering \cite{9}. In the field of electronics and materials, nanoelectronics for IoE, RF modules and packaging, high-frequency materials, radio transceivers, energy harvesting, THz imaging, and 2D/3D imaging will be involved. While in the field of computer science and engineering, the experts can contribute in various fields including image and signal analysis, mobile applications, security and privacy, big data analysis, smart sensor analytics, smart environment, ubiquitous systems. Experts active in the field of wireless communication play the main goal in the other fields including RF and antennas, 5G baseband, IoE application, future radio access, network optimization and management, spectrum regulations and channel modeling, etc.  

\section{Conclusion}
In this paper, we investigated how AI can contribute to the next generation of wireless communications. As a result, we conclude that 6G is not considered a faster 5G, but it is a convergence of a variety of technologies in various fields including communications, automation, control, sensing, and intelligence, which will be driven in context-aware, on-demand, self-configuration, and self-aggregation. By the introduction of 6G, we will experience life beyond rate-centric eMBB that is AI-powered communications (AIC), ultra-reliable and low latency communication (URLLC), super-eMBB, and massive connectivity. As a result, it is no exaggeration to claim that the next generation of mobile devices and network industry will be revolutionized by and built around the promise of AI.

\section*{Acknowledgment}

This work was supported by the Institute for Information \& communications Technology Promotion (IITP) grant funded by the Korea government (MSIP) (2018-0-01364, Terrestrial UHD based disaster broadcasting service for reducing disaster damage).

\vspace{12pt}

\end{document}